\documentclass[12pt,epsf,amssymb]{article}
\usepackage{epsfig,psfrag}
\usepackage{graphicx}
\usepackage{amsfonts}
\usepackage{amsmath}
\usepackage[usenames]{color}
\usepackage{amssymb}

\def\C{\mathbb{C}}

\begin{document}
\baselineskip 0.6cm
\newcommand{\gsim}{ \mathop{}_{\textstyle \sim}^{\textstyle >} }
\newcommand{\lsim}{ \mathop{}_{\textstyle \sim}^{\textstyle 3<} }
\newcommand{\vev}[1]{ \left\langle {#1} \right\rangle }
\newcommand{\bra}[1]{ \langle {#1} | }
\newcommand{\ket}[1]{ | {#1} \rangle }
\newcommand{\Dsl}{\mbox{\ooalign{\hfil/\hfil\crcr$D$}}}
\newcommand{\nequiv}{\mbox{\ooalign{\hfil/\hfil\crcr$\equiv$}}}
\newcommand{\nsupset}{\mbox{\ooalign{\hfil/\hfil\crcr$\supset$}}}
\newcommand{\nni}{\mbox{\ooalign{\hfil/\hfil\crcr$\ni$}}}
\newcommand{\EV}{ {\rm eV} }
\newcommand{\KEV}{ {\rm keV} }
\newcommand{\MEV}{ {\rm MeV} }
\newcommand{\GEV}{ {\rm GeV} }
\newcommand{\TEV}{ {\rm TeV} }

\def\diag{\mathop{\rm diag}\nolimits}
\def\tr{\mathop{\rm tr}}

\def\Spin{\mathop{\rm Spin}}
\def\SO{\mathop{\rm SO}}
\def\O{\mathop{\rm O}}
\def\SU{\mathop{\rm SU}}
\def\U{\mathop{\rm U}}
\def\Sp{\mathop{\rm Sp}}
\def\SL{\mathop{\rm SL}}

\def\change#1#2{{\color{blue}#1}{\color{red} #2}\color{black}\hbox{}}


\begin{titlepage}

\begin{flushright}
LTH/750
\end{flushright}

\vskip 2cm
\begin{center}
{\large \bf Metastable Vacua and Complex Deformations}
\vskip 1.2cm
Radu Tatar and Ben Wetenhall

\vskip 0.4cm

{\it Division of Theoretical Physics, Department of Mathematical Sciences

The University of Liverpool,
Liverpool,~L69 3BX, England, U.K.

rtatar@liverpool.ac.uk,~~benweten@liv.ac.uk}

\vskip 1.5cm

\abstract{We use the non-normalizable complex deformations to describe the stringy realizations of the metastable vacua in
${\cal N} = 1,~SU(N_c)$ SUSY theories with $N_f > N_c$ massive 
fundamental flavors. 
The consideration of the non-normalizable deformations requires a modified Toric duality. The new approach considers the 
tachyon condensation between pairs of wrapped D5 branes and anti D5 branes and the resulting mixing between some cycles in the geometry. We enlarge
 the class of  metastable vacua to the case of branes-antibranes wrapped on cycles of deformed $A_n$ singularities.   
} 

\end{center}
\end{titlepage}


\section{Introduction}

Over the last year there has been a lot of interest in studying some novel aspects of ${\cal N}=1$ SUSY field theories. Even though these theories have been 
the subject of intensive study in the 80's and 90's, there are some issues whose study started with the paper of Intriligator-Shih-Seiberg
\cite{iss}. The new idea is to consider a number of massive fundamental flavor greater than the number of colors and to go to the 
Seiberg dual theories where the F-term equations do not have any SUSY solutions. This implies that the theory is non-SUSY but it has been shown to be 
metastable. 

The paper \cite{iss} was followed by a substantial effort to get a string theory picture for the metastable vacua \cite{oo2}-cite{tw}.
In type IIA brane configurations with NS branes and color D4 branes and flavor given by D4 branes or D6 branes, the new vacua were claimed 
to be obtained by simply tilting the D4 branes on the directions of the NS branes. Unfortunately this picture has problems when lifted to M-theory
where the NS branes and D4 branes become a unique M5 brane \cite{bena}. The problem appears due to the inconsistency of the various asymptotic conditions
of the M5 branes. 

In the paper \cite{giku2007}, a scenario was proposed to take care of the objections of \cite{bena} by considering that 
the changes in the asymptotic conditions were  obtained after an infinite time. In their approach the Seiberg dual contains D4 branes and 
anti D4 branes with one end on the same NS brane and the other end on different orthogonal NS' branes. 
The tachyon condensation occurs on the NS brane, this determines a bending of the D4 branes and there is an
back reaction reaching the NS' branes in infinite time. 

In this work we present a general recipe to be used for dealing with the models with D5 branes wrapped on $P^1$ cycles of deformed ADE singularities.
They are dual to D4 on intervals between NS branes so the difficulties of IIA metastable brane configurations should be translated to IIB after 
a T-duality. Before going any further, we make a clear distinction between two type of geometries:

$\bullet$ deformations of the resolved $A_1$ singularity. 
By wrapping $N$ D5 branes on the single $P^1$ cycle which resolved the $A_1$ singularity, we get an ${\cal N} = 2, SU(N)$ theory. The field theory  has 
an adjoint field $\Phi$. By adding a superpotential $W(\Phi)$ of $m+1$-th power in $\Phi$ such that 
\begin{equation}
W'(\Phi) = g_{m+1}\prod_{i=1}^{m} (\Phi - a_i)
\end{equation}
we get an ${\cal N} = 1, \prod_{i=1}^{m} SU(N_i)$ with D5 branes wrapped on $m~~P^1$ cycles located at $\Phi = a_i$. All the $m~~P^1$ cycles are in the 
same homology class. 

One can also wrap anti D5 branes and get a non-SUSY system as proposed in \cite{agava1}. The D5-branes and anti D5
branes can annihilate each other by passing the barrier determined by the separation of the two $P^1$ in the same homology class. The T-dual for this 
was discussed in \cite{tw,marsano} and Landscape considerations were proposed in \cite{douglas}.

$\bullet$ deformations of the resolved $A_n$ singularities. 
By wrapping $N_i$ D5 branes on the $n~~P^1$ cycles which resolve the $A_n$ singularity, we get an ${\cal N} = 2, \prod_{i=1}^{n} SU(N_i)$ theory
with bifundamentals between neighbour groups. We add a quadratic superpotential for all the adjoint fields of the group $\prod_{i=1}^{n} SU(N_i)$. 
There are several considerations concerning the deformations:
  
1) To smoothen out the resulting geometry we need $n^2$ deformations. 

2) $n(n+1)/2$ of the deformations are $P^1$ cycles which can be pulled through a geometric transitions and correspond to normalizable deformations of 
the geometry after the geometric transition. 

3) $n(n-1)/2$ deformations do not correspond to $P^1$ cycles but to $S^3$ cycles  which are non-normalizable in the geometry after geometric 
transition. They measure the distance between various $P^1$ cycles. 

For a manifold with a certain intersection number between the $P^1$ cycles, the 
change in the intersection number is related to a change of complex structure. This change in complex structure is exactly realized by turning on the
non-normalizable complex deformations. 

What about the field theory? The distances between the various $P^1$ cycles correspond to the masses or vacuum expectation values for the 
bifundamental fields. The masses or vevs are functions of the coefficients in the tree level superpotential deforming the ${\cal N} = 2$ theory so 
the coefficients of the superpotential fix the sizes of the non-normalizable deformations. This is in accordance with the original 
Dijkgraaf-Vafa conjecture relating field theories and matrix theories \cite{dv1}. In their approach, the tree level superpotential is introduced in 
the exponential of a matrix model partition 
function and the result of the integration is the effective superpotential. 

In the paper \cite{dv2}, the bifundamental fields were considered to be massless and they only contributed with a factor to the measure of the partition 
function. In terms of geometry, this means that in their work the non-normalizable deformations are not turned on. In the cases considered in this work, 
the non-normalizable deformations are turned on and they determine some modifications of the usual Dijkgraaf-Vafa discussion. 
It would be interesting to consider this issue in
detail. In 
\cite{lm} it was also pointed out that the significance of the deformations exceeding the log-normalizable bound in Dijkgraaf-Vafa curve is less clear than the
normalizable ones. 

The non-normalizable deformations are displacements of the flavor cycles with respect to color cycles on the common 
direction of the normal bundle. For $N_f > N_c$, one goes to the Seiberg dual where the configurations contain branes and antibranes.
The non-normalizable deformations are forced to change by the brane-anti brane
pair annihilation  which determine them to combine with some of the normalizable deformations. 

Our argument go as follows:

$\bullet$ the geometric Seiberg dual is a flop or a Toric duality for toric manifolds. This holds nicely for massless flavors with zero vev. In that case,
even though we have D5 branes and anti D5 branes in the magnetic theory, they are on top of each other and 
the tachyon condensation ensuring a stable system does not change the geometry.

$\bullet$ change the complex structure of the electric theory geometry by changing the intersection number. After Seiberg duality, the result is a
magnetic theory with displaced D5 branes and anti D5 branes. The tachyon condensation determines the closure of some of the non-normalizable 
cycles. Because the D5 branes are wrapped on normalizable cycles, this can be obtained by combining a change of  some of 
the normalizable cycles which 
can be understood as a recombination of some of the  normalizable and 
non-normalizable cycles. The geometry now contains $P^1$ cycles which are not changed and they are holomorphic embedded in the geometry 
and cycles which have been combined with non-normalizable cycles. They are non-holomorphic in the original geometry but are metastable cycles.

It would be interesting to study these configurations as SUGRA solutions. This would imply to study D5 branes wrapped on 2-cycles 
obtained from the original $P^1$ cycle by deforming them
in the normal bundle directions and this would generalize the results of \cite{bdt1} - \cite{bdt6} where SUGRA solutions for D5 on $P^1$ cycle were built. 

We can also compare our results to the ones of \cite{agava1}. In their case there is also a clear distinction between the normalizable deformations
and non-normalizable deformations:

$\bullet$ The total number of deformations is $2 m - 1$, where $m$ is the highest power in $W'(\Phi)$. 

$\bullet$ The number of normalizable deformations is $m$ and the number of non-normalizable deformations is $m-1$. 

The main difference between our case and the one of \cite{agava1} is:

- in \cite{agava1} they considered the case of $A_1$ singularity deformed by a general superpotential. This way one obtains many $P^1$ cycles in the same
homology class which are separated by non-normalizable deformations. By putting D5 branes and anti D5 branes on cycles, they can cancel each other but the
$h_{11}=1$ is not modified and there is no recombination of cycles. As discussed in \cite{dot4}, the brane configuration involves a straight NS brane and a 
curved NS brane along $W'(x)$ where $W(x)$ is the tree level superpotential.

The intersection occurs at $W'(x) = 0$ and the D4 branes are located at these points in the direction $x$. The tachyon condensation occurs first on the 
straight NS brane and the stacks of D4 and anti D4 branes are forced to intersect on the straight NS brane. The bending of the D4 branes then propagates to 
the curved NS brane and the ends of the (anti) D4 branes on the curved NS brane can also touch each other by 
increasing the energy and moving on the curved NS brane. When the ends on the curved NS brane coincide,
there is total annihilation and the vacuum is without any D4 brane in type IIA or D5 branes in type IIB. 

- the situation is different for the deformation of $A_n$ singularity. The D5 branes and anti D5 branes are wrapped on cycle in different homology 
classes separated by non-normalizable  deformations. Due to the tachyon between  the D5 branes and anti D5 branes, there is a recombination of the cycles.  
In type IIA brane configurations, there are D4 branes and anti D4 branes 
with one end on the same NS brane and the other hand on two different NS branes. The 
phenomenon is as discussed in \cite{giku2007}, the ends of D4 and anti D4 branes on the same NS brane are connected after tachyon condensation.
Because the other end is on different NS branes, there is no total annihilation but the cycles are recombining in order to minimize the action 
of the branes. 

The normalizable cycles in some homology classes are unchanged and the normalizable cycles in other homology classes combine with the non-normalizable  
and become non-holomorphic embedded in the geometry.

\subsection{Comparison with Dynamical SUSY breaking with Fractional Branes}

The incompatibility between SUSY preservation, complex deformations and wrapped branes has been observed in \cite{hafr} for the case of 
toric varieties \footnote{We would like to thank Ami Hanany for pointing this to us.}.In their case, the complex deformations
smoothen out the singularities  and the fractional branes do not have any more vanishing cycles to wrap upon so they wrap finite cycles and their 
tension increases.

For the case of the suspended pinch point singularity 
\begin{equation}
x y = v w^2
\end{equation}
there are several types of fractional branes. Some of them trigger a complex deformation which provide a smooth surface whereas the 
${\cal N} = 2$ fractional branes only smoothen out a $x y = w^2,~~A_1$ part of the singularity parametrized by $z$ . In the presence of both types of 
fractional branes, the  ${\cal N} = 2$ fractional branes wrap finite cycles, get additional tension and break SUSY. The only way to 
recover SUSY is if the   ${\cal N} = 2$ fractional branes run to infinity.

In our case the role of the  ${\cal N} = 2$ fractional branes is played in the Seiberg dual theory by the flavor branes which hang 
between parallel NS branes so
only smoothen out a $A_1$ singularity. There is a tachyon condensation between these flavor branes and anti branes wrapped on 
other cycles and the result is metastable non-SUSY configurations. We can recover the SUSY if the    flavor branes are taken to infinity and the tachyon 
is removed. 
This corresponds to taking very large masses for the flavors which are then integrated out to reach a pure gauge theory. As the models we consider 
have light flavors, we always consider flavor branes at small distances from color branes so SUSY is broken. In section 5 we will make further comments and 
compare our  models with the elliptic models of \cite{argurio}. 

In section 2 we begin by reviewing the non-normalizable deformations. In subsequent sections we discuss several models where the understanding of the 
non-normalizable cycles is crucial for the description of metastable vacua.

\section{Complex Structure Deformations}
\label{sec:BottomUp}

We begin our discussion with a review of the results of \cite{dot3} concerning the classification of the 
complex structure deformation for a singular Calabi-Yau manifold. We will spell out the identification between 
the geometrical quantities and the field theory quantities. In the section 2.2 we will see how the non-normalizable deformations are 
related to the metastable non-SUSY vacua and SUSY vacua.

We start from the ${\cal N} = 2, A_{n-1}$ geometry 
\begin{equation}
x y = u^{n}
\end{equation}

We then consider the deformation to the case of the ${\cal N}=1$ geometry given by
\begin{equation}
\label{singular} 
F(x,y,u, v) :=xy - \prod_{k=0}^n \left(u- \sum_{i=0}^k g_i v^m\right)
=0, 
\end{equation} 
where $g_0$ is defined to be zero.

This admits the $U(1)$ symmetry group 
\begin{equation} 
x \rightarrow e^{i\theta/2} x,~~y \rightarrow
e^{i\theta/2}y,~~u \rightarrow e^{i\theta/(n+1)}u,~~v \rightarrow e^{i\theta/m(n+1)} v.
\end{equation} 
The miniversal
deformation space of the singularity, which describes the most general complex deformations, is given
by the chiral ring  
\begin{equation} 
\frac {\C \{ x, y, u,v \}}{\left(\partial F/\partial x,
\partial F/\partial y,\partial F/\partial u,\partial F/\partial v\right)}, 
\end{equation}
The dimension of the chiral ring is $n(nm+m-1)$.

For the case $m=1$ (only mass terms for the ${\cal N} = 2$ adjoint field), we get 4 deformations for the deformed $A_2$ singularity 
and 9 deformations for the $A_3$ singularity. For the deformation of an $A_1$ singularity with a cubic potential, the dimension of the 
chiral ring is 3.
 
The chiral ring characterises the geometry of the generic
deformation of $F =0.$  In order for the deformation to correspond to the dynamical 
parts of the theory at the singularity, the cohomology classes created by the deformation should be 
supported on the singularity i.e.
vanishing cohomologies.  The Poincar\'e duals to the vanishing cohomology classes are the vanishing homologies
which arises from quadratic singularities (i.e. conifold singularities). These classes correspond to the
normalizable (including log-normalizable) three forms.  This normalizability is necessary in order for
the deformation to describe the large N dual of the original theory because the original theory should be valid
near the singularity, and also insures that the geometric transitions will be the conifold transitions locally.

The normalizable deformations are generated by the monomials $u^iv^j$ with $mi +j \leq mn-1$ and there
are $m(n+1)n/2$ (3 for $m=1, n=2$; 6 for $m=1,n=3$; 2 for $m=2,n=3$) of them which give the dynamical parts of the dual theory 
(gluino condensation). 

There are $\frac{m(n+1)n}{2} - n$ (1 for $m=1,n=2$) non-normalizable
deformations which must be used to specify the theory externally (i.e. fixing the parameters of the tree-level
superpotential). So in the Milnor fiber  there are two kinds of $S^3$, ones corresponding to
the normalizable deformations and the others corresponding to the non-normalizable deformations.

Some particular cases of the general discussion are:

$\bullet$ $A_2$ quiver deformed by quadratic superpotential. We have 4 complex deformations, out of which 3 are normalizable deformation which  
correspond to gluino condensates and 1 non-normalizable which corresponds to parameters of the tree level 
superpotential. 

At the level of field theory, this case has been discussed in \cite{hanany-brodie} as 
an ${\cal N} =2, SU(N_1) \times SU(N_2)$ SUSY broken to ${\cal N}=1$ by adding the masses 
$m_1$ and $m_2$ for the two adjoint fields. The  ${\cal N}=1$ theory has an $ SU(\tilde{N}_1) \times 
SU(\tilde{N}_2)  \times SU(\tilde{N}_3)$ gauge group. The bifundamentals of the initial $SU(N_1) \times SU(N_2)$
theory become now  ${\cal N}=1$ fields which are either massive with a mass $\mu$ or have a vacuum expectation
value 
\begin{equation}
\label{def} 
\nu = \frac{\mu}{\frac{1}{m_1}+\frac{1}{m_2}}
\end{equation}
Let us consider the case $\tilde{N}_3=0$ and denote $\tilde{N}_1=N_c$ and $\tilde{N}_2 = N_f$. This is the case
with a gauge group $SU(N_c) \times SU(N_f)$ with massive bifundamentals. If we then take the coupling constant of 
$SU(N_f)$ to zero, this will provide us the case of an  ${\cal N}=1$ theory with $SU(N_c)$ gauge group and 
$N_f$ massive flavors with a mass $\mu$. 

In terms of the  geometrical deformations, the 3 normalizable complex deformations would correspond to the 
3 possible gluino condensates of the $SU(\tilde{N}_i),i=1,2,3$ groups. The non-normalizable deformation is measured 
by the value of $\nu$. The values of $m_i$ are very large, we see that this 
implies a very small value for $\mu$. Therefore the condition of \cite{iss} (very small value for the mass) is 
satisfied by the geometry.   

$\bullet$ $A_3$ quiver deformed by quadratic superpotential. We have 9 complex deformations, out of which 6 are normalizable deformation which would 
correspond to gluino condensates and 3 non-normalizable which correspond to parameters of the tree level 
superpotential. The field theory obtained on the 3 resolution $P^1$ cycle is $SU(N_1) \times SU(N_2) \times SU(N_3)$ with 
two pairs of bifundamental fields. The 3 non-normalizable can be either masses or expectation values for the bifundamental fields.

$\bullet$ $A_1$ quiver deformed by cubic superpotential. We have 3 complex deformations out of which 2 are normalizable deformations  which could 
correspond to gluino condensates and 1 is non-normalizable which corresponds to parameters of the tree level 
superpotential.

\subsection{Seiberg Dualities and Geometry Deformations}
\label{sec:Angles}

Consider the $SU(N_c)$ theory with $N_f > N_c$ flavors $Q, \tilde{Q}$. The Seiberg dual is $SU(N_f - N_c)$ with $N_f$ dual  flavors $q, \tilde{q}$,  
a gauge singlet (in the adjoint representation of the flavor group $SU(N_f)$), $M$ and a superpotential $q M \tilde{q}$. The electric flavors 
can be either massive or have expectation values. The corresponding situation in the dual theory is the following:

$\bullet$ give an expectation value to $n$ flavors. This breaks the gauge group to $SU(N_c - n)$ with $N_f - n$ fundamental flavors. The Seiberg dual is
$SU(N_f - N_c)$ with $N_f-n$ dual  flavors $q, \tilde{q}$ and  a gauge singlet $M$  (in the adjoint representation of the flavor group $SU(N_f-n)$).  
The gauge singlet loses components by getting an expectation value and the superpotential $q M \tilde{q}$ which implies that $n$ of the initial 
$N_f$ flavors in the dual theory acquire a mass.

$\bullet$ give a mass to $n$ flavors. If the mass is bigger that the dynamical scale, then the flavors are integrated out and the theory becomes 
 $SU(N_c)$ theory with $N_f - n$ flavors. If $N_f - n > N_c$, the Seiberg dual still exists but now it is $SU(N_f - N_c - n)$  
with $N_f-n$ dual  flavors $q, \tilde{q}$ and a gauge singlet in the adjoint representation of the flavor group $SU(N_f-n)$. The rest of $n$ flavors 
have acquired a vacuum expectation value  and this determines a linear term for some components of the singlet M.
If we give mass to more than $N_f - N_c$ then the dual group is completely Higgsed. 
We can  give mass to all the flavors and keep the mass under the dynamical scale so the flavor are light.

The geometrical interpretation for the Seiberg duality is as a flop in the geometry. The $SU(N_c)$ color group is obtained by wrapping D5 on a finite
$P^1$ cycle whereas the $SU(N_f)$ group is obtained by wrapping $N_f$ D5 branes on a holomorphic noncompact 2-cycle. These two cycles have a common
direction in their normal bundle. There are some bifundamentals in the group $SU(N_c) \times SU(N_f)$ which are actually the fundamental flavors if the 
group $SU(N_f)$ is a flavor group. If the two cycles touch, the bifundamentals are massless and have zero expectation value. If there is a displacement
between the two cycles on the normal bundle directions, then the bifundamentals have either mass or expectation value. 

The masses and expectation values of the  bifundamentals are related to the non-normalizable deformations of the 
geometry. The number of these  deformations should be the same in both electric and magnetic theory. The embedding of the color cycle $P^1$ into its normal
bundle imply the existence of two complex lines orthogonal to it. The two complex lines touch the  cycle $P^1$ at its North Pole and South Pole respectively.

When the Seiberg duality is seen as a flop in the geometry, the two complex lines are interchanged under a Seiberg duality.
If we choose to associate masses with  the displacement on one of such lines and associate vacuum expectation values to the displacement on the other line, 
then the Seiberg duality changes this convention so the map of masses into vacuum expectation values appears quite natural in the geometry.  

The relation between the non-normalizable deformations and the expectation values for flavors (same as linear terms in the singlet) or masses for the 
flavors in the Seiberg dual theories is helpful to checking that we have all the terms in the superpotential. 

Let us consider the $A_3$ singularity 
deformed with quadratic superpotentials. The ${\cal N} = 2$ theory for wrapped D5 branes is $\prod_{i=1}^3 SU(N_i)$ with the bifundamentals
$Q_1, \tilde{Q}_1$ in the $(N_1, \bar{N}_2), (\bar{N}_1, N_2)$ and $Q_2, \tilde{Q}_2$ in the $(N_2, \bar{N}_3), (\bar{N}_3, N_2)$. 
If the middle 2-cycle is 
finite and the other ones are semi-infinite then the theory can be viewed as ${\cal N} = 2, SU(N_2)$ with $N_1 + N_3$ fundamental flavors. 
There are 3 non-normalizable deformations which can be associated to either

- mass or vacuum expectation values for the $Q_1, \tilde{Q}_1$ quarks. This requires the existence of the term $Q_1 M_{11} \tilde{Q}_1 + M_{11}$.

- mass or vacuum expectation values for the $Q_2, \tilde{Q}_2$ quarks. This requires the existence of the term $Q_2 M_{22} \tilde{Q}_2 + M_{22}$.

- vacuum expectation for the $Q_1 Q_2$. This requires the existence of the term $Q_1 M_{12} \tilde{Q}_2$.

These cases are obtained from the superpotential
\begin{equation}
a~Q_1 M_{11} \tilde{Q}_1 + b~Q_2 M_{22} \tilde{Q}_2 +  c~Q_1 M_{12} \tilde{Q}_2 + d~M_{11} + e~M_{22}
\end{equation}
where the coefficients are functions of the electric variables and dynamical scales.

We can also consider the case of $A_5$ singularity and consider the Seiberg duality only for the even modes as done in \cite{amaritijune}. 
For $A_5$ singularity deformed with quadratic superpotentials we have 10 non-normalizable deformations. The ${\cal N} = 2$ gauge group would be 
$\prod_{i=1}^{5} SU(N_i)$ with four pairs of bifundamental fields $Q_{i}, \tilde{Q}_{i}, i=1,\cdots,4$. There are 10 possible vacuum expectation values for
the bifundamental fields and their products $Q_i Q_{i+1}$, $Q_{i} Q_{i+1} Q_{i+2}$, $Q_1 Q_2 Q_3 Q_4$. 

If we take the Seiberg dual only for the second and fourth
gauge group, this reduces to a collection of two $A_3$ singularities deformed by quadratic superpotentials. Therefore only 6 of the non-normalizable 
deformations are visible i.e. the vacuum expectation values or masses for for $Q_1, Q_2, Q_3, Q_4$ and vacuum expectation values for $Q_1 Q_2, Q_3 Q_4$. 
The superpotential allowing these deformations is
\begin{equation}
Q_1 M_{11} \tilde{Q}_1 + Q_2 M_{22} \tilde{Q}_2 +  Q_3 M_{33} \tilde{Q}_3 + Q_4 M_{44} \tilde{Q}_4 +  
Q_1 M_{12} \tilde{Q}_2 + Q_3 M_{34} \tilde{Q}_4 + M_{11} + M_{33} + M_{55}  
\end{equation} 

The main question of this work is how to handle with these non-normalizable deformations for the models with metastable non-SUSY vacua. Considering 
that the Seiberg duality is a flop in the geometry for the case when the non-normalizable deformations are not present, what is the 
effect of turning on the latter? We consider the extra deformations  in the electric theory and see what is the corresponding quantity 
on the magnetic theory side. We will see that the non-normalizable deformations combine with some 
of the normalizable deformations to form new cycles.

\section{${\cal N} =1, SU(N_f) \times SU(N_c)$ Model}

Consider the ${\cal N} =2, SU(N_f) \times SU(N_c)$ model and deform it in two ways.
The deformation  ${\cal N} =2 \rightarrow  {\cal N} =1$ is
\begin{equation}
\label{potelewoutlin}
W_{ele} = \frac{1}{2} \tilde{\mu}  \tilde{\Phi}^2 +  \frac{1}{2} \mu  \Phi^2 +  \tilde{Q} (\Phi + \tilde{\Phi}) Q 
+ \xi \Phi +  \tilde{\xi} \tilde{\Phi}
\end{equation}
The F-term equations are
\begin{equation}
\label{1}
0 = \Phi Q + Q \tilde{\Phi},
\end{equation} 
\begin{equation}
\label{2}
0 = \tilde{Q} \Phi + \Phi \tilde{Q},
\end{equation}
\begin{equation}
\label{3}
0 = \mu \Phi + Q \tilde{Q} + \xi,
\end{equation}
\begin{equation}
\label{4}
0 = \tilde{\mu} \tilde{\Phi} + \tilde{Q} Q + \tilde{\xi}.
\end{equation}
As discussed in \cite{gipe}, there are several solutions of these equations:

1) The electric theory corresponds to $\mu \tilde{\mu} \ne 0$ when both adjoint fields are massive. 

If the linear terms are zero $\xi = \tilde{\xi} = 0$, the solution with $SU(N_f) \times SU(N_c)$ group (i.e. no
expectation value for the fundamental flavors) is $\Phi = \tilde{\Phi} = 0$.

If the linear terms are not  zero $\xi~\tilde{\xi} \ne 0$, the situation changes. $Q$ and $\tilde{Q}$ can be simultaneously diagonalized 
by a color rotation and they assume the form:
\begin{equation}
Q = \mbox{diag}(q_a),~~~~~~\tilde{Q} =  \mbox{diag}(\tilde{q}_a),
\end{equation}
the equations of motion implying that $Q \tilde{Q}$ have at most one non-vanishing eigenvalue. This eigenvalue is 
the displacement of the NS branes in the brane configurations or the displacement of the cycles in the geometry.
These displacements can correspond to either masses or vevs for the fundamental fields. 

2) The magnetic theory corresponds to $\mu \ne 0$ and  $\tilde{\mu} = 0$ when only one of the 
adjoint fields is massive. The other adjoint field becomes the meson singlet and remains massless. 

If the linear terms are zero $\xi = \tilde{\xi} = 0$, the expectation values for  $Q, \tilde{Q}$ are zero so 
$\tilde{\Phi}$ is free. The  $\tilde{\Phi}$ is the singlet meson field in the dual theory. 

If the linear term  $\tilde{\xi} \ne 0$, the paper \cite{gipe} considered the SUSY solution which required 
$N_c \ge N_f$. In the magnetic theory that we consider, the number of colors is $N_f - N_c$ and the number of 
massive flavors is $N_f$ so the SUSY is broken by the rank condition  the SUSY in broken by the  rank condition i.e. the 
F-term equation for the field $\tilde{\Phi}$ does not have a solution, same as in \cite{iss}. The SUSY breaking 
solution is similar to the one in \cite{iss}.

By continuously changing the parameters from the case 1) to case 2) we go from the electric to magnetic theory
by a Seiberg duality. We now want to describe the geometrical interpretation of this duality.  

\subsection{The corresponding Geometry}

In order to discuss the geometrical Seiberg duality for the model with light flavors, we start with a simpler one, the  
${\cal N} =1, SU(N_f) \times SU(N_c)$ theory with massless flavors. 

We are going to discuss the electric and magnetic brane configurations for this theory. The T-duality between the 
brane configurations and geometry implies that we need to add an extra $S^1$ circle to the D4 branes in order to get 
D5 branes on $S^1 \times \mbox{interval}$ which is the same as D5 branes on $P^1$ cycles. The NS branes are mapped into lines of
singularity living in the the normal bundle to the $P^1$ cycles. Even though we will only draw brane configurations, we understand that there is an extra
$S^1$ which is the extra dimension of the D5 branes \cite{dot1}-\cite{dot5}. 

In order to discuss the Seiberg duality we start with a resolved ${\cal N} = 2, A_3$ singularity. Each $P^1$ cycle has a normal 
bundle:
\begin{equation}
X' = X,~~Y' = Y Z^2,~~Z'=1/Z
\end{equation}
The  ${\cal N} = 2, A_3$ singularity is then deformed into a collection of resolved conifold singularities, each one looking like
\begin{equation}
X' = X Z,~~Y' = Y Z,~~Z'=1/Z
\end{equation}
The number of normalizable deformations is 6 and the number of the non-normalizable deformations is 3. If we close all the non-normalizable cycles, we remain with
3 normalizable $P^1$ cycles which touch each other. In Figure 1, we denote  the 3 $P^1$ cycles of the  ${\cal N} = 1$ resolved geometry by A, B and C from left to right. 
We always understand that the cycles A and C have very large volume as they are considered flavor cycles. 

The electric brane configurations involves $N_f$ D5 branes wrapped on A cycle and $N_c$ D5 branes wrapped on B cycles as in Figure 1.

\begin{figure}
\begin{center}
\setlength{\unitlength}{1mm}
\begin{picture}(80,40)  
\put(5,20){\line(1,0){50}}
\put(5,5){\line(0,1){30}}
\put(75,5){\line(0,1){30}}
\put(20,8){\line(1,1){30}}
\put(55,5){\line(0,1){30}}
\end{picture}
\caption{Electric configuration of branes with $N_f$ Massless Quarks}
\end{center}
\end{figure}
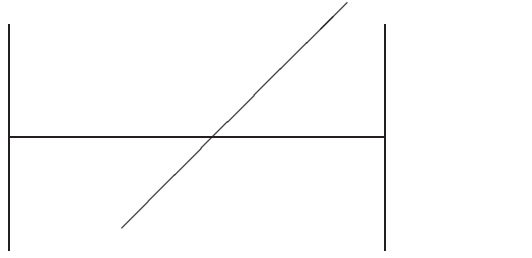

We now use the results of \cite{beasley,hehana,vafafi} which considered the Seiberg duality as a Toric duality. In our case, the 
Toric duality implies shrinking down the cycle A and blowing up the cycle C. This is because the geometry can separated into a 
resolved conifold singularity and a resolution of a deformed $A_2$ singularity. This can be done in two ways and 
the results of \cite{vafafi} imply that the 
Seiberg duality is a Weyl reflection in the Dynkin diagram of $A_3$. In terms of the cycles this means
\begin{equation}
A \rightarrow A' = A + B,~~B' \rightarrow - B.
\end{equation}
which satisfy the charge conservation condition
\begin{equation}
N_f A + N_c B = N_f A' + (N_f - N_c) B'.
\end{equation}  
The cycle A and C are interchanged  by a flop because they touch the B cycle at its North and South pole respectively, and the 
poles of the B cycle are interchanged. The resulting picture is  in Figure 2 where the order of cycles is now A, B and C from right to left.  
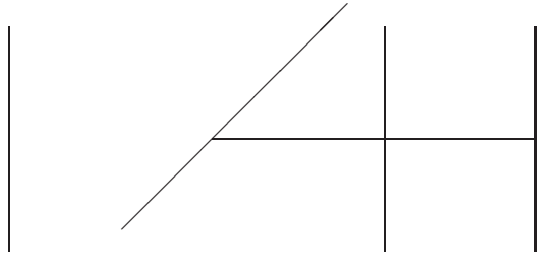
\begin{figure}
\begin{center}
\setlength{\unitlength}{1mm}
\begin{picture}(80,40)
\put(32,20){\line(1,0){43}}
\put(5,5){\line(0,1){30}}
\put(75,5){\line(0,1){30}}
\put(20,8){\line(1,1){30}}
\put(55,5){\line(0,1){30}}
\end{picture}
\caption{Magnetic configuration of branes with $N_f$ Massless Quarks}
\end{center}
\end{figure}

The $N_c$ D5 branes on the $B$ cycle changed into $N_c$ anti D5 branes on the $B'$ cycle and the $N_f$ D5 branes on the 
$A$ cycle changed into $N_f$ branes on the $A'$ cycle. 

There is a tachyon condensation between the   $N_c$ anti D5 branes on the $B'$ cycle and $N_c$ of the $N_f$ D5 branes. Because the 
D5 branes and the anti D5 branes are on top of each other, the cycles of the geometry are not modified by the tachyon condensation
The result is a configuration with $N_f - N_c$ D5 branes on the $B'$ cycle and $N_f$ D5 branes on the A cycle of the 
dual geometry.  

The discussion for the case of massive quarks is more involved. The electric geometry is Figure 3. To understand it we need to make some observations.

\begin{figure}
\begin{center}
\setlength{\unitlength}{1mm}
\begin{picture}(80,40)  
\put(5,15){\line(1,0){22}}
\put(32,20){\line(1,0){23}}
\put(5,5){\line(0,1){30}}
\put(75,5){\line(0,1){30}}
\put(20,8){\line(1,1){30}}
\put(55,5){\line(0,1){30}}
\end{picture}
\caption{Electric configuration of branes with $N_f$ Massive Quarks}
\end{center}
\end{figure}
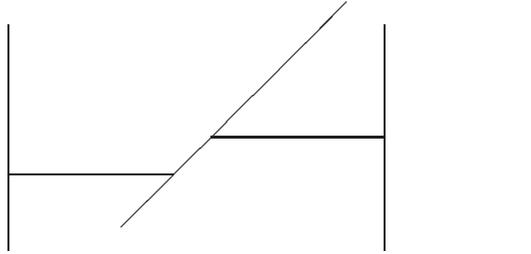

There are two ways to represent the fundamental matter for ${\cal N} =2, SU(N_f) \times SU(N_c)$ brane configuration. 

1) One way involves three parallel NS branes and two sets of $N_f$ D4 branes and $N_c$ D4 branes between them. 

2) The second way is with  two parallel NS branes together with $N_c$ D4 branes and $N_f$ orthogonal D6 branes.

To go to the ${\cal N} = 1$ theory, the rotation of one of the  NS branes is related to giving a mass $\mbox{tan}(\theta)$ to the adjoint field, where 
$\theta$ is the rotation angle. For the case 2), the rotation of the D6 branes is related to changing the Yukawa coupling by 
a factor $\mbox{cos}(\alpha)$ where $\alpha$ is the rotation angle of either D6 or both NS branes. 

In the configuration with only NS branes and D4 branes there are several ways to rotate the NS branes:

$\bullet$ the rotation of the middle or the rightmost NS brane by an angle 
$\theta$ implies a mass for the adjoint field equal to   $\mbox{tan}(\theta)$. 

$\bullet$ the rotation of the leftmost NS brane by an angle 
$\alpha$ implies a factor $\mbox{cos}(\alpha)$ in front of the Yukawa coupling. 

$\bullet$ by keeping the middle NS brane unchanged and rotating the outer NS branes by the same angle $\theta$, this gives a mass
 $\mbox{tan}(\theta)$ to the  ${\cal N} =2$ adjoint field and also puts a factor  $\mbox{cos}(\theta)$

We can now understand the difference between Figure 1 and Figure 3. We see that there no way to directly go from one to another
because the stacks of D4 branes are stuck because they lie between non parallel NS branes. The only way to deform one into another is to use the non-normalizable 
deformations of the theory. For an $A_2$ singularity there are 3 normalizable and 1 non-normalizable deformation. The latter 
is related to adding a linear term to the  ${\cal N} =2$ adjoint field which is exactly related to turning on mass
or vacuum expectation value for the fundamental quarks. 

What about the magnetic picture? As the Seiberg duality is a Toric duality, the magnetic picture should look as in Figure 4.

\begin{figure}
\begin{center}
\setlength{\unitlength}{1mm}
\begin{picture}(80,40)
\put(32,20){\line(1,0){23}}
\put(27,15){\line(1,0){48}}
\put(5,5){\line(0,1){30}}
\put(75,5){\line(0,1){30}}
\put(20,8){\line(1,1){30}}
\put(55,5){\line(0,1){30}}
\end{picture}
\caption{Magnetic configuration of branes from $N_f$ Massive Quarks}
\end{center}
\end{figure}
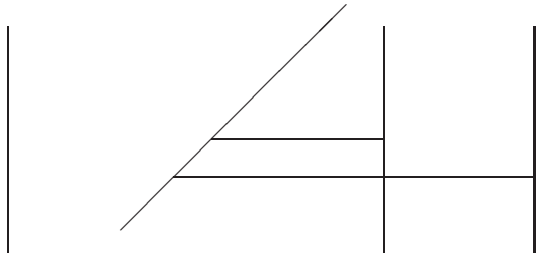

 Can this geometry also come from 
some non-normalizable deformation of an  ${\cal N} =2$ theory? The deformation for the  ${\cal N} =2, SU(N_f) \times SU(N_c)$
would be
\begin{equation}
\frac{m}{2} \Phi_1^2 + \xi_1 \Phi_1 + \xi_2 \Phi_2
\end{equation}
because there is no mass term for $\Phi_2$. $\Phi_2$ is exactly the meson singlet in the Seiberg dual theory. 

The effect of the first two terms in the superpotential is to create a displacement between the 
cycles $A'$ and $B'$ on the line of singularity corresponding to the right NS brane. Now let us put the corresponding 
$N_c$ anti D5 branes on the $B'$ cycle and $N_f$ D5 branes on the $A'$ cycle. 

The effect of the tachyon condensation is more 
complicated than in the massless case because it has to undue the non-normalizable deformation. 
As argued in \cite{giku2007}, the condensation occur first on the left NS brane and then propagates to the right NS branes as a 
kink. The authors of \cite{giku2007} have calculated the bending of the D4 branes on the directions of NS branes. In the geometric
picture, this maps into a bending of the holomorphic $P^1$ cycle in its normal bundle directions. 

The bending of the cycle with $N_c$ D5 branes is T-dual to the bending of the $x^6$ direction of the D4 brane into 
the $x^5$ direction discussed in \cite{giku2007}. Their DBI equation of motion argument gave the result for bending as
\begin{equation}
\delta{x} = \frac{1}{2~l} (y_1^2 \mbox{sin} 2 \theta_1 +y_2^2 \mbox{sin} 2 \theta_2)
+ l (\theta_1 + \theta_2)
\end{equation}
where $l$ is the characteristic scale, $x$ is the displacement of the $P^1$ cycle in the direction of the non-normalizable cycle,
$y$ is the size of the two cycles and $\mbox{cos} \theta_i = \frac{y_m}{y_i}$ for 
$y_m$ the minimum of $y$.  In the cases considered in \cite{giku2007}, this solution was quite complicated and involved several types of transitions.

The simplification in our case is the fact that the position of the rightmost NS brane is at infinity because the 
corresponding cycle is semi infinite as being a flavor cycle. The mass of the fundamental flavors is also small so 
the $x$ displacement is small. As discussed in \cite{giku2007}, in this case the dependence of 
$x$ on $y$ is monotonic and we do not have any other maxima or minima. Therefore, the tachyon condensation in 
our case will give two stacks of D5 branes, one with D5 branes $N_f - N_c$  wrapped on the $A'$ cycle and 
one with $N_c$ D5 branes wrapped on the deformation of the $A$ cycle after the flop.

One needs to consider the extra complications which appear when T-dualizing non-SUSY configurations, as discussed in 
\cite{marsano} for deformations of $A_1$ geometry. It would be interesting to generalize their discussion to the 
case of $A_n$ for general $n$.

We can also make connection to the M-theory discussion. The T-duality between the wrapped D5 branes and Hanany-Witten configurations
followed by a lift of the latter to M theory allows a direct identification between the cycles of the cycles in IIB geometry and
cycles of the Riemann surface which is part of the M5 brane. The M5 brane has 1-cycles corresponding to the non-normalizable cycles 
in IIB. By closing the non-normalizable cycle, there are cycles in the Riemann surface which remain holomorphic and some which 
are not holomorphic. The non-normalizable 
deformation makes the difference between the two types of cycles as first discussed in \cite{franco,bena}.

\section{${\cal N} =1, SU(N_f-N_c) \times SU(N_c) \times SU(N_c)$ Model}

In this section we consider the situation discussed in \cite{oodec2006}. Before doing so, we make some comments on the 
relation between configurations with flavors given by D6 branes and the ones with flavors given by D5 branes wrapped on 
semi infinite 2-cycles. 

Consider first the configuration with NS branes and D6 branes. The ${\cal N} =2$ model includes parallel NS and D6 branes
which are orthogonal to the NS branes. 
Let us now split  the  single stack of D6 branes into one stack containing $n_1$ D6 branes and one stack containing $n_2$ D6 branes. 
We then rotate the right NS brane
by an angle $\theta$ and the stack of $n_1$  D6 branes by the same angle $\theta$. As discussed in \cite{giku}, there is a 
repulsive force between D6 branes which are not parallel so the rotation breaks the flavor group $SU(n_1+n_2)$ 
into $SU(n_1) \times SU(n_2)$. Denote the $n_1$ quarks by $Q_1, \tilde{Q}_1$ and the  $n_1$ quarks by $Q_2, \tilde{Q}_2$.
The superpotential contains the terms
\begin{equation}
\mbox{cos} \theta Q_1 \Phi \tilde{Q}_1 +   \mbox{cos} \theta Q_2 \Phi \tilde{Q}_2 + \frac{1}{2} \mbox{tan}{\theta} \Phi^2
\end{equation}
By integrating out the adjoint field we get the quadratic coupling
\begin{equation}
 Q_1 \tilde{Q}_1  Q_2 \tilde{Q}_2 
\end{equation}
We then take the stack of $n_1$  D6 branes to the right infinity and  the stack of $n_2$  D6 branes to the left infinity.
There is a brane creation Hanany-Witten effect which implies that the  $n_1$  D6 branes are connected by $n_1$ D4 branes 
to the left NS brane and  $n_1$  D6 branes are connected by $n_2$ D4 branes to the right NS brane. 

We can then exchange the model with D6 brane flavors with  one with D4 branes between NS branes on semi infinite intervals. 
The T-dual of the latter is 
represented by replacing the  D4 branes with D5 branes wrapped on non compact cycles and the NS branes with the normal bundle to the 
$P^1$ cycles. In the   ${\cal N} =2$ model, all the four NS branes are parallel. 
In the rotated model, the rotation of the $n_1$  D6 branes by an angle $\theta$ is equivalent to the rotation of the rightmost NS by the same angle
$\theta$. The ${\cal N} =1$ model discussed above has the first and the third NS branes parallel to each other and rotated by an 
angle $\theta$ with respect to  the second and fourth NS branes which are parallel and unrotated. 

The unrotated IIB geometry is  ${\cal N} = 2, A_3$. We can solve the singularity and wrap branes on the cycles to get 
${\cal N} =2, SU(N_f-N_c) \times SU(N_c) \times SU(N_c)$ model. The two above rotations imply that all the  adjoint fields will 
get a mass equal to $\mu = \mbox{tan} \theta$. There is also a change in the coupling between the bifundamentals and the 
adjoint fields , which acquire an extra factor $y = \mbox{cos}~\theta$. We can also add a linear term in the 
adjoint fields of type $ \xi_i \Phi_i$. The superpotential becomes
\begin{equation}
\label{potele1}
W_{ele} = \sum_{i=1}^3 \frac{1}{2} \mu  \Phi_i^2 + 
y \tilde{Q}_1 (\Phi_1 + \Phi_2) Q_1 +  y \tilde{Q}_2 (\Phi_2 + \Phi_3) Q_2
+ \sum_{i=1}^{3} \xi_i \Phi_i
\end{equation}
The F-term equations are
\begin{equation}
\label{11}
0 = \Phi_1 Q_1 + Q_1 \Phi_2,~~0 = \tilde{Q}_1 \Phi_1 + \Phi_2 \tilde{Q}_1
\end{equation}
\begin{equation} 
\label{21}
0 = \Phi_2 Q_2 + Q_2 \Phi_3,~~0 = \tilde{Q}_2 \Phi_2 + \Phi_3 \tilde{Q}_2
\end{equation}
\begin{equation}
\label{31}
0 = \mu \Phi_1 + y Q_1 \tilde{Q}_1 + \xi_1,
\end{equation}
\begin{equation}
\label{51}
0 = \mu \Phi_2 + y \tilde{Q}_1 Q_1 + Q_2 \tilde{Q}_2 + \xi_2,
\end{equation}
\begin{equation}
0 = \mu \Phi_3 + y \tilde{Q}_2 Q_2 + \xi_3
\end{equation}

If we put $\xi_2 = 0$, we see that, besides the quartic term in the superpotential, we also get the mass terms
\begin{equation}
\frac{\xi_1}{\mu}  Q_1 \tilde{Q}_1 +  \frac{\xi_2}{\mu}  Q_2 \tilde{Q}_2
\end{equation}

In the geometry, the values of $\xi_1, \xi_2$ are exactly the sizes of the non-normalizable cycles in the 
geometry. From the general formula, we see that for the deformations of the $A_3$ singularity with quadratic superpotentials 
for $\Phi$  we have 3 non-normalizable deformations,  two of them being $\xi_1$ and $\xi_2$.

The solution of the F-term equations implies that the bifundamentals become massive 
with masses $\xi_1/\mu$,~$\xi_2/\mu$. In terms of the geometry we have three 2-cycles denoted from left to right by A, B, C. 
There are $N_f - N_c$ D5 branes wrapped on the A cycle, $N_c$ on the B cycle and $N_c$ on the C cycle.
As discussed before, the displacement of the A cycle with respect to the B cycle in the common normal bundle direction is 
equal to the size of the non-normalizable deformation $\xi_1$ and the displacement of the C cycle with respect to the B cycle in 
the common normal bundle direction is equal to the size of the non-normalizable deformation $\xi_2$.

We can now perform the Seiberg duality as a flop in the geometry or as a Weyl reflection in the $A_3$ algebra. As before, we 
interchange the A and C cycle and the change in cycles is
\begin{equation}
A \rightarrow A' = A + B, B' \rightarrow - B, C \rightarrow C' = C + B
\end{equation}
 
In the Seiberg dual, there are $N_f - N_c$ D5 branes wrapped on $A'$ cycle, $N_c$ anti D5 branes on the $B'$ cycle and 
$N_c$ D5 branes on the $C'$ cycle. 
There are tachyons between the D5 branes and anti D5 branes. There should be a tachyon condensation in 
order to obtain a supersymmetric configuration. If $\xi_1 > \xi_2$, the condensation is between the branes wrapped on the 
$B', C'$ cycles.  If $\xi_1 < \xi_2$, the condensation is between the branes wrapped on the $A', B'$ cycles.

If we consider the case of  \cite{oodec2006}, $\xi_1 > \xi_2$ and the condensation appears between the $N_c$ anti D5 branes on the 
$B'$ cycle and $N_c$ D5 branes on the $C'$ cycle. The result is that the  $N_f - N_c$ D5 branes wrapped on $A'$ cycle remain 
unchanged and the other 5-branes are wrapped on a cycle bent in the $x'$ direction where the direction  $x'$ is obtained 
from $x$ after a rotation by an angle $\theta$. The dependence of $x'$ on $y$ is again monotonic due to 
the fact that the right most NS brane is at infinity.

\section{Relation to Elliptic Model Metastable Vacua}

We now discuss the relation of the non-normalizable deformations to the deformation of the metastable elliptic model considered in \cite{argurio}. 
Their geometry, being a deformation of the orbifolded conifold, also come from an ${\cal N} =2$ theory so it makes sense to distinguish between 
normalizable and non-normalizable deformations. 

We start by making an observation about the deformation of the orbifolded conifold. In \cite{ot1}, the deformations of the 
orbifolded conifold have been considered. For the $Z_k \times Z_l$ orbifold it has been shown that the singular manifold is  
smoothen out by 
\begin{equation}
kl+k+l-2~~~\mbox{deformations}.
\end{equation}
Out of this, 

$\bullet$ $k+l-2$ come from separating $k l$ singular conifold points 

$\bullet$ $k l$ come from the resolution of these conifold singularities. 

The $k l$ deformations are cycles which can go through geometric transitions and are normalizable cycles whereas the 
$k+l-2$ are related to the tree level superpotential and are non-normalizable deformations. 

Because we want to consider the case of the $Z_3$ orbifold of the 
conifold, we take $k=3, l=1$ which implies that we have 2 non-normalizable deformations  coming from separating the conifold points
and 3 normalizable deformations from the actual resolution of the conifold singularities. These two types of deformations are exactly what 
was called in \cite{franco1} as  three fractional branes deformations and two ${\cal N} =2$ fractional branes deformation. 

We can also describe a map between the (non)normalizable deformations of elliptic and non-elliptic models. 
The corresponding non-elliptic model is an $A_1$ singularity with a quartic potential for the adjoint field. 

The deformations of the  $k=3, l=1$ orbifolded conifold are mapped into this deformations of the $A_1$ singularity. 
The latter has 5 deformation out of which 3 are normalizable (corresponding to 
gluino condensates) and 2 are non-normalizable (corresponding to vacuum expectation values or linear terms in the 
superpotentials). 

Then one maps the  3 deformation fractional branes into the  3 normalizable deformations and the 2 ${\cal N} =2$ fractional 
brane deformations into the  2 non-normalizable deformations. 

By making this identification, it results that the ${\cal N} =2$ fractional 
brane deformations correspond to masses or vacuum expectation values for quarks. 
One can map the general case of \cite{agava1} to any elliptic model coming from orbifolding the conifold 
with arbitrary $k$ and $l=1$.

\section*{Acknowledgements}

We would like to dedicate this work to the memory of our good friend and collaborator Kyungho Oh. The work of Radu Tatar is funded by PPARC and 
thanks the Galileo Galilei Institute for Theoretical Physics and the INFN for partial support during the 
completion of this work. RT also thanks Benasque Center for Physics for hospitality during the last stages of this work.

\end{document}